\documentclass[osa,twocolumn,showpacs]{revtex4}

\usepackage{here}
\usepackage[dvips]{graphicx}

\newcommand\pictc[5]{\begin{figure}
                       \centerline{
                       \includegraphics[width=#1\columnwidth]{#3}}
                   \protect\caption{\protect\label{fig:#4} #5}
                    \end{figure}            }
\newcommand\pict[4][1.]{\pictc{#1}{!tb}{#2}{#3}{#4}}
\newcommand\rpict[1]{\ref{fig:#1}}

\newcommand\leqt[1]{\protect\label{eq:#1}}
\newcommand\reqtn[1]{\ref{eq:#1}}
\newcommand\reqt[1]{(\reqtn{#1})}

\newcounter{Fig}

\begin{document}
\begin{sloppy}

\title{Soliton X-junctions with controllable transmission}

\author{Andrey A. Sukhorukov}
\homepage{http://www.rsphysse.anu.edu.au/nonlinear/ans}
\affiliation{Nonlinear Physics Group,
Research School of Physical Sciences and Engineering,
Australian National University,
Canberra, ACT 0200, Australia}

\author{Nail N. Akhmediev}
\homepage{http://www.rsphysse.anu.edu.au/~nna124/}
\affiliation{Optical Sciences Centre,
Research School of Physical Sciences and Engineering,
Australian National University,
Canberra, ACT 0200, Australia}

\begin{abstract}
We propose new  planar X-junctions and multi-port devices written by spatial solitons, which are composed of two (or more) nonlinearly coupled components in Kerr-type media. Such devices have no radiation losses at a given wavelength. We demonstrate that, for the same relative angle between the channels of the X-junctions, one can {\em vary the transmission coefficients} into the output channels by adjusting the {\em polarizations of multi-component solitons}. 
We determine analytically the transmission properties and suggest two types
of experimental embodiments of the proposed device.
\end{abstract}

\maketitle

\pict{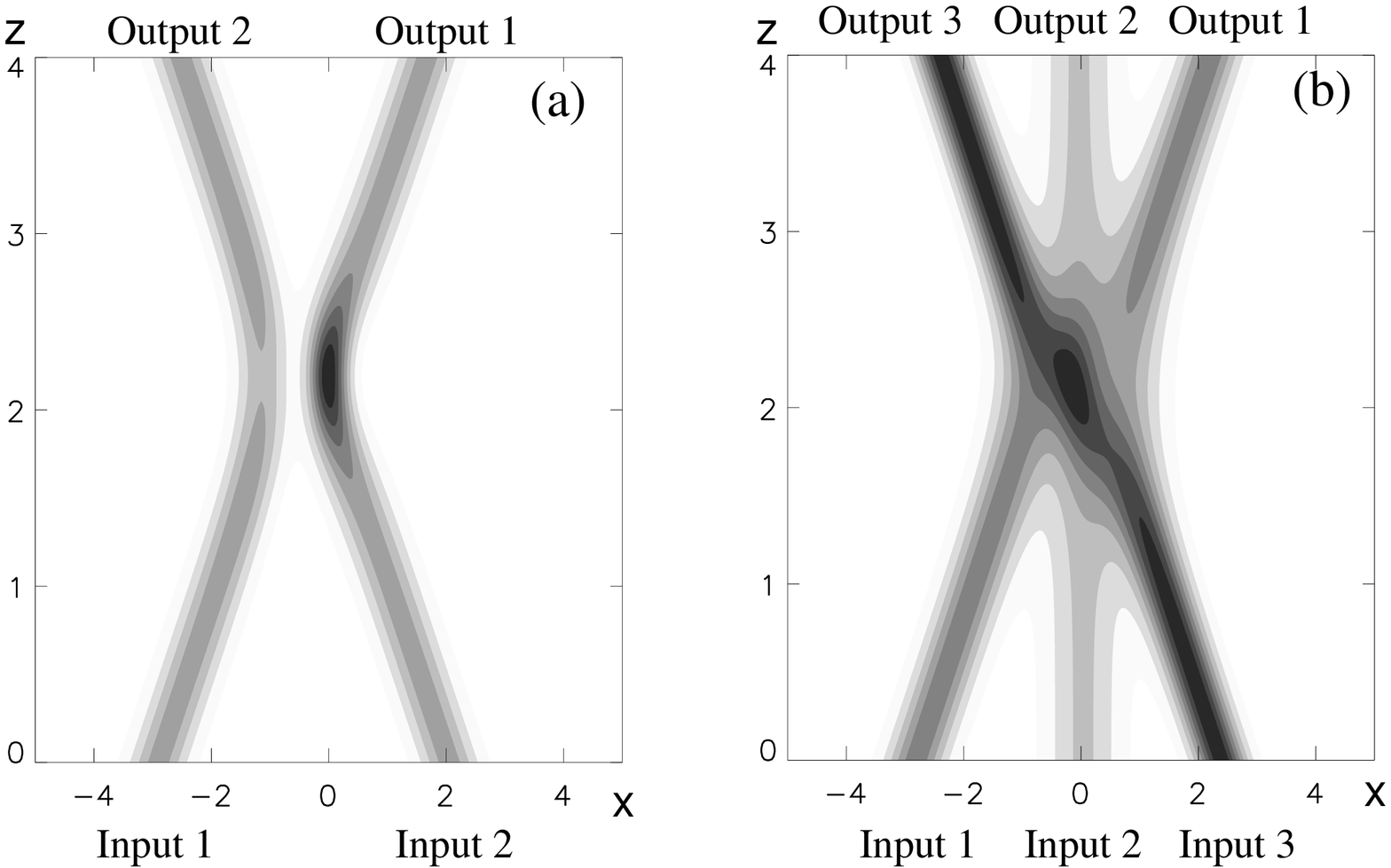}{index-two}{
Spatial distribution of refractive index created by (a)~two colliding multi-component solitons with amplitudes $r_1 = r_2 = 2$, velocities $\mu_1 = -\mu_2 = 1$, and relative polarization angle $\tan^{-1}(1/\rho_1^-) = \pi /3$, and (b)~three orthogonally polarized solitons with amplitudes $r_1=2$, $r_2=1.5$, $r_3=2.5$, and velocities $\mu_1=-\mu_3=1$, $\mu_2=0$.
}

Optical X-junctions and couplers are devices that have two waveguide channels with a weak interaction between them, so that they can swap signals after some distance~\cite{Snyder:1983:OpticalWaveguide}. 
In more general terms an X-junction or a coupler is a four-port device with two inputs and two outputs arranged so that the output signals depend on the parameters of the input signals (e.g., wavelength) and some internal controlling parameters (e.g., coupling coefficients). 

It was suggested that two colliding spatial solitons, that are formed in a nonlinear Kerr-type medium, induce a change of refractive index, and this can be viewed as a linear X-junction or coupler~\cite{Akhmediev:1993-186:OC}, see Fig.~\rpict{index-two}(a). Similarly, multi-port devices can be formed by several colliding solitons, see see Fig.~\rpict{index-two}(b).
Such a waveguide device can be permanently set after its creation in a photorefractive medium~\cite{Guo:2001-1274:OL}.
The transmission characteristics of the coupler depend on the soliton intensities, and on the angle of incidence between solitons~\cite{Miller:1996-4098:PRE}. On the other hand, the radiation losses are completely absent at a given wavelength. In this work, we extend the analysis of soliton-written devices to the case of vectorial, or multi-component, interactions. This extension allows us to increase significantly the flexibility of the device as the transmission coefficients become multi-parameter functions of the angle between the channels and the polarization angles. In previous works, only one of these parameters has been used.

We consider the interaction dynamics of ``multi-component solitons'', which are composed of two or more incoherently coupled components in a medium with Kerr-type self-focusing nonlinearity. Such solitons can be observed, for example, in photorefractive crystals~\cite{Christodoulides:1996-1763:APL, DelRe:2001-61:SpatialOptical}. We note that the soliton polarization state is not directly related to the orientation of the electric field, but rather characterized the power distribution between several mutually incoherent components, which are combined into a soliton.
The soliton evolution along the propagation direction can be modeled, in the parabolic approximation, by a system of nonlinear Schr\"odinger equations for the set of modes, which are coupled through the change of refractive index. In the case of (1+1)-D spatial geometry, the normalized equations are
\begin{equation} \label{eq:nls}
  i \frac{\partial \psi_{m}}{\partial z}
  + \frac{1}{2} \frac{\partial^{2} \psi_{m}}{\partial x^{2}}
  + I(x,z) \psi_m
  = 0 ,
\end{equation}
where $\psi_m$ is the normalized amplitude of the $m$-th component ($1 \le m \le M$), $M$ is the number of components, $z$ is the coordinate along the direction of propagation, and $x$ is the transverse coordinate. We assume that the change of refractive index is proportional to the total intensity,
\begin{equation} \leqt{I}
   I(x,z) = \sum_{m=1}^M |\psi_m(x,z)|^2 . 
\end{equation}
Such an approximation is valid, in particular, for the description of photorefractive screening solitons in the low-saturation regime.

It was demonstrated that Eqs.~(\ref{eq:nls}) are {\em completely integrable} by means of the inverse scattering technique (IST)~\cite{Manakov:1973-505:ZETF, Makhankov:1982-55:TMF} and, therefore, any localized solution can be represented as a nonlinear superposition of solitary waves. Specifically, the solution for $N$ interacting bright solitons can be found by solving a set of auxiliary linear equations~\cite{Nogami:1976-251:PLA}, which we present in the following form,
\begin{equation} \leqt{slau}
   \sum_{n^\prime=1}^{N}\sum_{m^\prime=1}^{M} 
        u_{n^\prime m^\prime}
        \frac{\delta_{m^\prime m}
              + (\chi_{n m} e_n) 
                (\chi_{n^\prime m^\prime} e_n^\prime)^\ast}{
              k_n +k_{n^\prime}^{\ast}}
    = - \chi_{n m} e_n .
\end{equation}
Here $e_n = \exp \left( \beta_n + i \gamma_n \right)$, $\beta_n = r_n (x - \mu_n z)$, $\gamma_n = \mu_n x + (r_n^2-\mu_n^2) z / 2$, $k_n = r_n + i \mu_n$ are the complex soliton wavenumbers ($r_n>0$ with no loss of generality), and the complex numbers $\chi_{n m}$ are arbitrary constants. The $N$-soliton solution of the original Eq.~\reqt{nls} is then obtained by adding up of all the $u_{n m}$ functions corresponding to a given component number $m$,
\begin{equation} \leqt{psi}
   \psi_m(x,z) = \sum_{n=1}^{N} u_{n m}(x,z) .
\end{equation}
Finally, a linear soliton-written device is characterized by the refractive index profile that is fixed according to Eqs.~\reqt{I} and~\reqt{psi}. As a matter of fact, the functions $u_{n m}$ form {\em a full set of localized modes} of the corresponding linear Eq.~\reqt{nls}, if the laser frequency is not changed. Below, we use this property to calculate the transmission properties of soliton-written devices.

We consider the situation when the input and output coupler channels are formed by well separated solitons. Then, solutions of Eqs.~\reqt{slau} can be expressed as decompositions over the profiles of individual, non-interacting, solitons,
\begin{equation} \leqt{Unm}
   u_{n m}^\pm = \sum_{j=1}^{N} 
                   U_{n m j}^\pm 
                     {\rm sech}\left\{ r_j [x -x_{j}^\pm 
                                   - \mu_j (z - z^\pm)] \right\}
                     e^{i \gamma_{j}} ,
\end{equation}
where the signs $-$ and $+$ correspond to the input (at $z=z^-$) and output (at $z=z^+$) channels, respectively, and $x_n^\pm$ define the soliton positions. 

In order to find the complex coefficients $U_{n m j}^\pm$, we analyse the properties of original Eqs.~\reqt{slau}. First, we note that only $N$ modes are linearly independent, since
\begin{equation} \leqt{depend}
   \chi_{n m_2} \sum_{n^\prime=1}^{N} 
                 \frac{u_{n^\prime m_1}}{k_n + k_{n^\prime}^\ast}
   = 
   \chi_{n m_1} \sum_{n^\prime=1}^{N} 
                 \frac{u_{n^\prime m_2}}{k_n + k_{n^\prime}^\ast} ,
\end{equation}
for arbitrary $n$, $m_1$, and $m_2$. Second, we use the fact that for well separated solitons $|e_n| \gg |e_{n^\prime}|$, if $x_n \ll x_{n^\prime}$, and obtain the following relations:
\begin{equation} \leqt{nosingul}
   \begin{array}{l} {\displaystyle
      S_\chi(n, j) \equiv
      \sum_{m^\prime=1}^{M} U_{n m^\prime j }^\pm 
                            (\chi_{n m^\prime})^\ast = 0,
      \,\, x_n^\pm \ll x_{j}^\pm , \qquad
   } \\*[9pt] {\displaystyle
      S_k^\pm(n, m, j) \equiv
      \sum_{n^\prime=1}^{N} \frac{U_{n^\prime m j}^\pm}{
                             k_{n}+k_{n^\prime}^\ast} = 0,
      \,\, x_n^\pm \gg x_{j}^\pm ,
   } \\*[9pt] {\displaystyle
      x_j^\pm = \mu z^\pm + \ln( C_j^\pm) / r_j , \,\,
      C_j^\pm = \sqrt{ \frac{S_k^\pm(j, m, j)}{
                       \chi_{j m} S_A^\pm(j, j) S_\chi^\pm(j, j) }},\qquad
   } \\*[9pt] {\displaystyle
      2 S_k^\pm(j, m, j) + C_j^\pm \left[1 + S_B^\pm(j, j)\right] = 0 ,
   } \end{array}      
\end{equation}
where 
\begin{equation} \leqt{SAB}
   \begin{array}{l} {\displaystyle
      S_{A}^\pm(n, j) \equiv
      \sum_{n^\prime=1}^{N} \frac{A_{n^\prime j}^\pm}{
                             k_{n}+k_{n^\prime}^\ast} = 0, \,\,
                              x_n^\pm \ll x_{j}^\pm , \qquad
   } \\*[9pt] {\displaystyle
      S_{B}^\pm(n, j) \equiv
      \sum_{n^\prime=1}^{N} \frac{B_{n^\prime j}^\pm}{
                             k_{n}+k_{n^\prime}^\ast} = -1, \,\,
                             x_n^\pm \ll x_{j}^\pm , 
   } \\*[9pt] {\displaystyle
      A_{n j}^\pm = B_{n j}^\pm = 0, \,\,
                              x_n^\pm \gg x_{j}^\pm , \qquad 
      A_{j j}^\pm = 1, \,\, B_{j j}^\pm = 0 .
   } \end{array}      
\end{equation}
Using these equations, one can calculate all the complex amplitudes $U_{n m j}^\pm$, and the soliton shifts that are expressed through the constants $C_j^\pm$. Additionally, the input and output polarizations of multi-component solitons, can be found with the help of Eqs.~\reqt{psi} and \reqt{Unm}. We note that in earlier studies the polarization rotation was only calculated for two-soliton collisions~\cite{Radhakrishnan:1997-2213:PRE, Kanna:2001-5043:PRL}.

\pict{fig02.eps}{transfer}{
Power splitting between the two output channels vs. the relative polarization angle $\tan^{-1}(1/\rho_1^-)$ of two multi-component solitons, with the wavenumbers being the same as in Fig.~\rpict{index-two}(a).}

A multi-port linear device can be conveniently characterized by the complex transmission matrix ${\bf T}$. By definition, 
\begin{equation} \leqt{transfer}
   V_j^+ = \sum_{j^\prime=1}^{N} {\bf T}_{j j^\prime} V_{j^\prime}^- ,
\end{equation}
where $V_j^\pm$ are the mode amplitudes at the input ($-$) and output($+$) channels created by the soliton number $j$. On the other hand, according to Eq.~\reqt{Unm}, $V_j^\pm = U_{n m j}^\pm$ should satisfy the relation~\reqt{transfer} for arbitrary $n$ and $m$, and these conditions can be used to uniquely determine the transmission matrix coefficients. 

We now apply the above general analytical results to the description of an X-junction, which is created by two colliding multi-component solitons ($N=M=2$). An example of refractive index distribution in such planar device is shown in Fig.~\rpict{index-two}(a). When only the first input channel is illuminated, i.e. $V_2^- = 0$, then the power in the output channels can be found in a simple analytical form. Using notations introduced in Ref.~\cite{Jakubowski:1998-6752:PRE}, we can write the resulting expressions as follows,
\begin{equation} \leqt{power2s}
   \begin{array}{l} {\displaystyle 
      \frac{P_1^+}{P_1^-} 
         \equiv |T_{1 1}|^2 
         = \frac{1 + |\rho_1^-|^2}{1+|\rho_1^+|^2}, \quad
   } \\*[9pt] {\displaystyle 
      \frac{P_2^+}{P_1^-}
         \equiv \frac{r_2}{r_1} |T_{2 1}|^2
         = \frac{r_2}{r_1} \frac{1 + |\rho_1^-|^2}{1+|\rho_2^+|^2} ,
   } \end{array}
\end{equation}
where 
\begin{equation} \leqt{rho}
   \begin{array}{l} {\displaystyle 
      \rho_1^- = \frac{\eta_2 \eta_1 + 1}{\eta_2 - \eta_1},\quad
      \rho_1^+ = \rho_1^- \frac{k_1+k_2^\ast}{k_1 - k_2},\quad
   } \\*[9pt] {\displaystyle 
      \rho_2^+ = \frac{1}{(\rho_1^-)^\ast} 
                 \left[ (k_1+k_2^\ast) 
                        \frac{1 + |\rho_1^-|^2}{2 r_1} - 1 \right],
   } \end{array}
\end{equation}
and $\eta_j$ characterize the input polarizations of multi-component solitons, for which $|\psi_1 / \psi_2| = \eta_j$, where $j=1,2$ is the soliton number. 
It follows from Eqs.~\reqt{power2s} and~\reqt{rho} that the {\em power splitting can be tuned by changing the relative polarizations} of multi-component solitons at the input, while the soliton amplitudes and propagation angles remain the same, as illustrated in Fig.~\rpict{transfer}. The {\em maximum cross-talk} is observed in the case of scalar solitons. On the other hand, solitons with orthogonal polarizations create {\em zero cross-talk} intersections. By tuning the relative soliton polarization, it is possible to obtain any desired power splitting in between the two extremes. For example, the junction shown in Fig.~\rpict{index-two}(a) has a 1:1 splitting ratio, despite the fact that the spatial distribution of the refractive index is asymmetric. The properties of multi-port devices can be engineered in a similar way. For example, multiple colliding solitons having orthogonal polarizations induce zero cross-talk intersections, such as shown in Fig.~\rpict{index-two}(b).

Our analytical results suggest the possibility of the following experimental realizations. We consider two different arrangements.
{\em 1. ``Light-guiding-light'' type of experiment.} This can be done in a photo-refractive medium with the two soliton beams crossing each other at a fixed angle. Each of the beams contains two mutually incoherent components, and the power distribution between these component determines the soliton polarization state. Suppose the polarization of one of the beams is fixed and the polarization of the other one can vary. Then controlling this angle allows to change the transmission coefficients of the probe beam launched onto any of the channels. 
{\em 2. Fixed X-junction experiment.} The waveguide can be prepared in advance in a photo-sensitive medium using photolithography or equivalent technique. Computer assisted equipment allows us to make multi-port devices using exact analytical expressions. For example, transmission coefficient of an X-junction can be made equal to any value shown in Fig.~\rpict{transfer} for a {\em fixed angle} between the channels. The latter can be important in the design of optical integrated circuits where the choice of angles is restricted by the design features.

In conclusion, we have demonstrated that the X-junctions formed by colliding multi-component solitons in Kerr-type media can be used as ideal single-level optical integrated circuits, that can share the signal in any desired proportion between output channels. We also found that it is possible to construct zero cross-talk waveguide intersections. Such devices can be designed to have no radiation losses at any given wavelength. The parameters of arbitrary soliton induced multi-port devices can be optimized using solutions of coupled algebraic equations. The latter can readily be obtained numerically using our analytic results. NA acknowledges support from the US Army Research Office - FE. The work of AS has been partially supported by the Australian Research Council.

\end{sloppy}
\end{document}